\documentclass[12pt]{iopart}
\usepackage{graphicx}
\usepackage{esdiff}
\usepackage{subfig}
\usepackage{comment}
\usepackage{hhline}
\usepackage{array}
\usepackage{textcomp}
\usepackage{hyphsubst}
\usepackage{hyperref}
\usepackage{upgreek}
\usepackage[dvipsnames]{xcolor}

\begin{document}

% \title[Numerical model of microwave SQUID multiplexer resonator characteristics]{Numerical model of microwave SQUID multiplexer resonator characteristics with power dependence and junction inhomogeneity}
\title[Advanced microwave SQUID multiplexer model] {Advanced microwave SQUID multiplexer model incorporating readout power effects and Josephson junction inhomogeneities}

\author{M Neidig$^{1}$, M Wegner$^{2,1}$ and S Kempf$^{1,2}$}
\address{$^1$ Institute of Micro- and Nanoelectronic Systems, Karlsruhe Institute of Technology, Hertzstrasse 16, Building 06.41, D-76187 Karlsruhe, Germany}
\address{$^2$ Institute for Data Processing and Electronics, Karlsruhe Institute of Technology, Hermann-von-Helmholtz-Platz 1, Building 242, D-76344 Eggenstein-Leopoldshafen}
\ead{\mailto{martin.neidig@kit.edu}}
\vspace{10pt}
\begin{indented}
\item[]December 2025
\end{indented}

\begin{abstract}
We present an advanced  model for describing the readout power dependence of the resonance characteristics of a microwave SQUID multiplexer. Our model proves valid for SQUID screening parameters up to $\beta_\mathrm{L}<1$, hence covering the full range of practically relevant design parameters. We demonstrate that our model significantly improves agreement with experimental data compared to the existing models, thereby enabling optimization beyond the previously accessible parameter space. Moreover, our model supports non-sinusoidal current-phase relations of the rf-SQUID's Josephson junction, allowing, for the first time, for the modeling of devices based on Josephson tunnel junctions with inhomogeneous tunnel barriers. We show that the effects of such inhomogeneities are qualitatively similar to, yet distinct from, those of the screening parameter, making their inclusion essential for accurate characterization. Incorporating these effects yields great improved agreement with measurements, even at readout power conditions well beyond typical operating parameters.
\end{abstract}

\vspace{2pc}

\maketitle

\section{Introduction}

Cryogenic microcalorimeters, such as superconducting transition-edge sensors (TESs)~\cite{Irw05, Ull15} and magnetic microcalorimeters (MMCs)~\cite{Fle05, Kem18}, provide exceptional energy resolution for detecting incident particles. Although latest advances in modern micro- and nanofabrication technologies enable the realization of TES and MMC arrays of virtually any size, the development of suitable SQUID-based multiplexing schemes has lagged behind, thereby limiting the practically achievable number of readout channels. 
Kinetic inductance current sensors (KICS)\cite{Szy24} have also emerged as a promising alternative readout technique, but this technology is not yet sufficiently mature for widespread deployment in large detector arrays.
At present, microwave SQUID multiplexing ($\upmu$MUX)~\cite{Irw04, Mat08, Kem17b} represents the most promising approach for scaling TES and MMC arrays beyond $10^3$ individual detectors, owing to its high multiplexing factor, low on-chip power dissipation, and wide bandwidth per readout channel.
However, as the $\upmu$MUX contributes to and often even sets the overall noise performance of the entire detector system, optimizing its design and operating parameters is essential. This optimization requires accurate models that reliably predict $\upmu$MUX behavior and enable precise device characterization. 
Wegner~{\it et~al.}\cite{Weg22} introduced an analytical $\upmu$MUX model that accurately describes the dependence of the characteristics of state-of-the-art $\upmu$MUX devices on the microwave probe tone power and reproduces their measured behavior under typical operating conditions. As such, it constitutes a core component of simulation frameworks used for device optimization~\cite{Sch23, Red24}. However, because the model is based on Taylor series expansions, its validity is limited to SQUID screening parameters $\beta_\mathrm{L} = 2 \pi L_\mathrm{S} I_\mathrm{c}/ \Phi_0 < 0.6$. Here, $L_\mathrm{S}$, $I_\mathrm{c}$, and $\Phi_0$ denote the SQUID inductance, critical current of the Josephson junction, and magnetic flux quantum, respectively. This constraint restricts the applicable parameter space, and consequently that of simulations relying on it, even though a wider parameter range is of practical interest for modern device design. To overcome this inherent limitation, we have developed a numerical simulation framework that accurately models the power dependence of $\upmu$MUX devices up to $\beta_\mathrm{L} < 1$, beyond which a $\upmu$MUX gets hysteretic and changes operational behavior.

In this paper, we present the structure of our simulation framework and compare its results with those obtained from the analytical model and experimental data. Furthermore, the framework supports arbitrary current–phase relations of the Josephson tunnel junction, enabling, for the first time, the modeling of $\upmu$MUX devices comprising junctions with inhomogeneous tunnel barrier thickness. We analyze the effects of such inhomogeneities in comparison with those arising from the screening parameter and demonstrate that including them in the $\upmu$MUX model significantly improves agreement with experimental observations.

% -----------------------------------------------

\section{Basics of microwave SQUID multiplexing}

In the following, we introduce the basic concepts of microwave SQUID multiplexing, which are described in more detail in \cite{Mat11, Kem17}. We define the nomenclature used throughout this paper and outline the origin of the $\upmu$MUX power dependence, as presented in \cite{Weg22}, up to the point where our advanced $\upmu$MUX model deviates from the existing ones.

\begin{figure}
    \centering
    \includegraphics[width=0.5\linewidth]{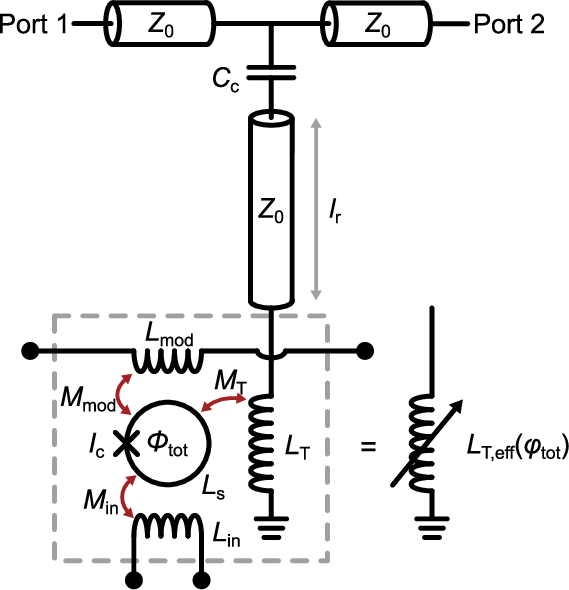}
    \caption{Equivalent circuit diagram of a single $\upmu$MUX channel. The channel comprises an rf-SQUID containing a Josephson junction with critical current $I_\mathrm{c}$ and a closed superconducting loop with inductance $L_\mathrm{S}$. The SQUID is inductively coupled to the input coil $L_\mathrm{in}$ and the inductor $L_\mathrm{T}$, which loads a superconducting quarter-wave microwave resonator. The resonator is coupled to a transmission line via a coupling capacitor $C_\mathrm{c}$. A modulation coil $L_\mathrm{mod}$ is coupled to the SQUID for flux ramp modulation. Owing to the mutual couplings, the load inductor can be described by an effective inductance $L_\mathrm{T,eff}(\varphi_\mathrm{tot})$ that depends on the total magnetic flux $\Phi_\mathrm{tot}$ through the SQUID, with $\varphi_\mathrm{tot}=2\pi\Phi_\mathrm{tot}/\Phi_0$.}
    \label{fig:mux_scheme}
\end{figure}

Figure~\ref{fig:mux_scheme} shows an equivalent circuit diagram of a single $\upmu$MUX readout channel. The detector signal is inductively coupled via the input coil $L_\mathrm{in}$ to an rf-SQUID, which is, in turn, coupled to the load inductor $L_\mathrm{T}$ of a superconducting quarter-wave microwave resonator. The opposite end of the resonator is capacitively coupled, through the coupling capacitor $C_\mathrm{c}$, to a transmission line that shares the same characteristic impedance $Z_0$ as the resonator\footnote{The characteristic impedance of the transmission line does not necessarily have to match that of the resonator. However, for simplicity, this assumption is adopted throughout this work.}.
The load inductor $L_\mathrm{T}$ can be represented by an effective inductance $L_\mathrm{T,eff}(\varphi_\mathrm{tot})$, which varies periodically with the normalized total magnetic flux $\varphi_\mathrm{tot}=2\pi\Phi_\mathrm{tot}/\Phi_0$ threading the SQUID loop.
Consequently, the resonance frequency $f_\mathrm{r}$ of the circuit gets dependent on $\varphi_\mathrm{tot}$ and can be expressed as
\begin{equation}
    f_\mathrm{r}(\varphi_\mathrm{tot}) = \frac{f_0}{1 + 4f_0(C_\mathrm{c}Z_0 + L_\mathrm{T,eff}(\varphi_\mathrm{tot})/Z_0)}.
    \label{eq:fr}
\end{equation}
Here, $f_0$ denotes the resonance frequency of the unloaded resonator, that is set by its physical length.

A detector signal alters the total magnetic flux threading the SQUID loop, thereby shifting the loaded resonance frequency $f_\mathrm{r}$. This shift is typically measured by applying a microwave probe tone at port 1 at a frequency close to resonance of the $\upmu$MUX channel, i.e., with an angular frequency of $\omega \approx \omega_\mathrm{r} = 2\pi f_\mathrm{r}$, and monitoring the transmitted microwave signal at port 2.
To linearize the nonlinear relationship between resonance frequency and input signal, flux ramp modulation \cite{Mat12} is employed, which requires an additional inductor $L_\mathrm{mod}$ inductively coupled to the rf-SQUID.

To accurately model the magnetic flux dependence of the effective load inductance $L_\mathrm{T,eff}(\varphi_\mathrm{tot})$, and thereby the resulting resonance shift, it is essential to consider the currents running through the load inductance $L_\mathrm{T}$. In this way, the effective load inductance can be expressed as
\begin{equation}
    L_\mathrm{T,eff}=L_\mathrm{T}\frac{i_\mathrm{tot}(t)}{i_\mathrm{T}(t)}=L_\mathrm{T}\left(1+\frac{i_\mathrm{ind}(t)}{i_\mathrm{T}(t)}\right),
    \label{eq:LT}
\end{equation}
where the total current through the load inductance, $i_\mathrm{tot}(t) = i_\mathrm{T}(t) + i_\mathrm{ind}(t)$, consists of two contributions: the current $i_\mathrm{T}(t)$ generated by the microwave probe tone and the induced current $i_\mathrm{ind}(t)$ resulting from the supercurrent in the rf-SQUID.
The current resulting from the microwave probe tone with angular frequency $\omega$ is given by 
\begin{equation}
    i_\mathrm{T}(t) = I_\mathrm{T}\sin{(\omega t)},
\end{equation}
where its amplitude
\begin{equation}
    I_\mathrm{T}=\sqrt{\frac{16}{\pi}\frac{Q_\mathrm{l}^2}{Q_\mathrm{c}}\frac{P_\mathrm{rf}}{Z_0}}
\end{equation}
depends on the on-chip microwave power $P_\mathrm{rf}$, as well as the loaded quality factor $Q_\mathrm{l}$ and the coupling quality factor $Q_\mathrm{c}$ of the resonator.
The current induced by the supercurrent inside the rf-SQUID is given by
\begin{equation}
    i_\mathrm{ind}(t)=-\frac{M_\mathrm{T}}{i\omega L_\mathrm{T}}\frac{\mathrm{d}I_\mathrm{S}(t)}{\mathrm{d}t},
    \label{eq:ind}
\end{equation}
where $M_\mathrm{T}$ denotes the mutual inductance between the load inductor and the SQUID loop, and $I_\mathrm{S}(t)$ is the supercurrent circulating in the SQUID.

Josephson tunnel junctions, the non-linear elements routinely employed in $\upmu$MUXs for parametric conversion, are typically assumed to possess homogeneous tunnel barriers with low transparency, leading to the well-known current-phase relation $I_\mathrm{S,JJ}(\varphi)=I_\mathrm{c}\sin{(\varphi)}$. Here, $I_\mathrm{c}$ is the junction's critical current and $\varphi$ is the phase difference across it. From this, the expression for the supercurrent circulating the rf-SQUID can be derived: 
\begin{eqnarray}
    I_\mathrm{S}(t)&=&-I_\mathrm{c}\sin{(\varphi_\mathrm{tot})} \nonumber\\
    &=&-I_\mathrm{c}\sin{\left[\varphi_\mathrm{ext}+\varphi_\mathrm{rf}\sin{( \omega t)}+\beta_\mathrm{L}\frac{I_\mathrm{S}(t)}{I_\mathrm{c}} \right]}.
    \label{eq:Is}
\end{eqnarray}
Here, we consider that the total normalized magnetic flux $\varphi_\mathrm{tot}$ threading the rf-SQUID loop comprises multiple components. The input and modulation signal, applied via the modulation and input coil, vary on timescales orders of magnitude slower than the microwave probe tone and can therefore be treated as a quasi-static external flux contribution $\varphi_\mathrm{ext}=2\pi\Phi_\mathrm{ext}/\Phi_0$. 
The microwave probe tone contributes a sinusoidal magnetic flux signal with amplitude $\varphi_\mathrm{rf}=2\pi\Phi_\mathrm{rf}/\Phi_0$ and $\Phi_\mathrm{rf}=M_\mathrm{T}I_\mathrm{T}$. Finally, the screening flux, caused by the current induced in the rf-SQUID, is given by $\varphi_\mathrm{scr}(t)=\beta_\mathrm{L}I_\mathrm{S}(t)/I_\mathrm{c}$, where $\beta_\mathrm{L}=2\pi L_\mathrm{S}I_\mathrm{c}/\Phi_0< 1$ is the screening parameter of the non-hysteretic rf-SQUID.

\begin{figure*}[t]
    \centering
    \includegraphics[width=1\linewidth]{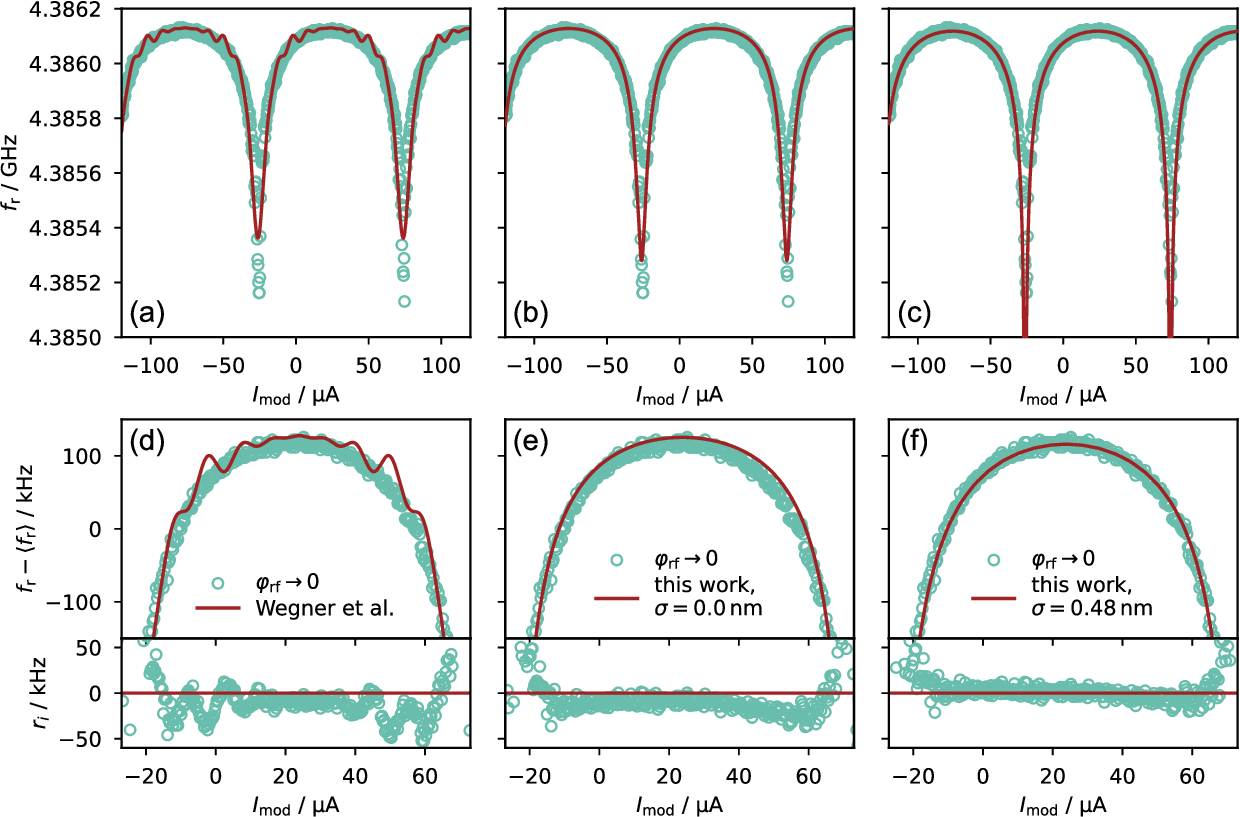}
    \caption{Measured dependence of the resonance frequency $f_\mathrm{r}(I_\mathrm{mod})$ on the modulation coil current $I_\mathrm{mod}$ for a representative $\upmu$MUX channel. The corresponding resonance curves were recorded using a vector network analyzer with low microwave probe tone power, corresponding to the limit $\varphi_\mathrm{rf} \rightarrow 0$. Panel (a) shows a fit to the data using the analytical model described by Wegner~{\it et.~al.}~\cite{Weg22}. In (b) the data was fitted using the model presented in this work, assuming an ideal (i.e., perfectly homogeneous) Josephson tunnel junction barrier. Panel (c) shows a fit obtained with our new model assuming an inhomogeneous tunneling barrier thickness with Gaussian probability density distribution characterized by a mean barrier thickness $\bar{d}=2\,\mathrm{nm}$ and standard deviation $\sigma=0.48\,\mathrm{nm}$). Panels (d), (e), and (f) show magnified views of the fits in the panels above, where $\langle f_\mathrm{r} \rangle = 4.386\,\mathrm{GHz}$ denotes the resonance frequency averaged over $I_\mathrm{mod}$. The residuals $r_i$ of the fits are plotted at the bottom of the figure, with the red line marking $r_i=0$. The best agreement between model and data is achieved using the model that includes a finite tunnel barrier inhomogeneity.}
    \label{fig:models_phirf0}
\end{figure*}

The main challenge in modeling the dependence of the resonance frequency shift on the external magnetic flux $\varphi_\mathrm{ext}$ and the microwave probe tone power $\varphi_\mathrm{rf}$ lies in solving the implicit equation~\ref{eq:Is}. Since there is no closed-form solution for $I_\mathrm{S}(t)$, the model presented by Wegner {\it et al.}~\cite{Weg22} employs an analytical, 10$^\mathrm{th}$-order Taylor expansion in $\beta_\mathrm{L}$ to approximate a solution. This analytical model shows excellent agreement with experimental data for screening parameters $\beta_\mathrm{L}<0.6$. However, due to the inherent limitations of the Taylor expansion, the model deviates from the expected behavior and exhibits unphysical features beyond its valid range. As an example, figures~\ref{fig:models_phirf0} (a), (d) show the resonance frequency shift of a $\upmu$MUX channel with $\beta_\mathrm{L}\approx 0.65$, measured with a probe tone power $\varphi_\mathrm{rf}\rightarrow 0$ and fitted with the analytical model. In fact, we fitted the analytical model, as well as the models introduced in the following sections, to data acquired at four different readout powers (see figure~\ref{fig:shift_multiple_Pwrs}), using maximum-likelihood fits \cite{iminuit}. Data taken at the lowest readout power were excluded for resonance frequencies below $4.3856,\mathrm{GHz}$, where the strong dependence of the resonance frequency on the externally applied magnetic flux leads to the largest uncertainties in our magnetically unshielded measurement setup.
The fit with the analytical model exhibits unphysical ripples that become even more pronounced as $\beta_\mathrm{L}$ increases. Although the model can, in principle, be extended to higher values of $\beta_\mathrm{L}$, this would require a substantially higher order in the Taylor expansion, rendering the approach in fact impractical.

% -----------------------------------------------

\section{Advanced µMUX modeling based on numerical simulation framework} 

\begin{figure}
    \centering
    \includegraphics[width=0.5\linewidth]{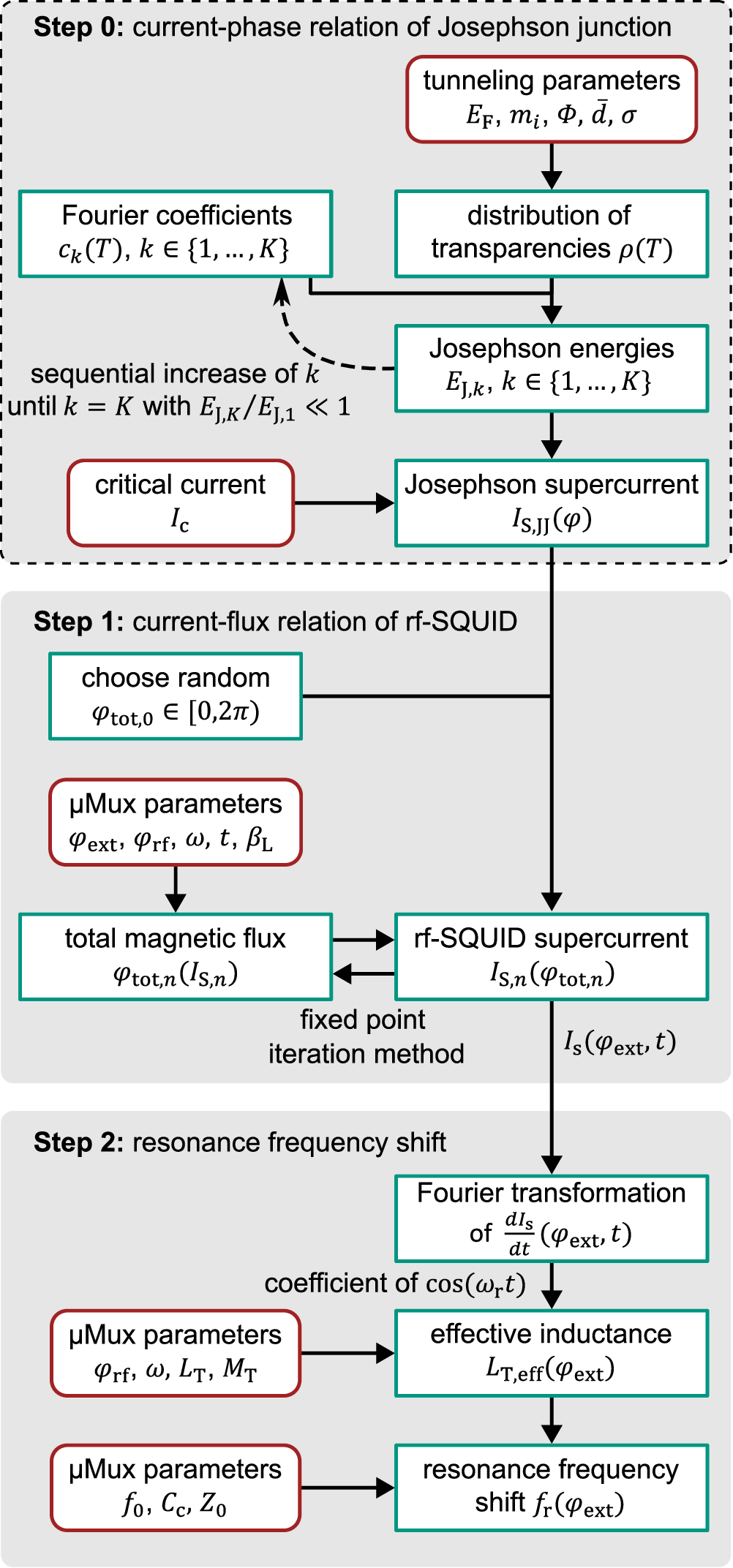}
    \caption{Flowchart outlining our numerical simulation framework. Red, rounded boxes indicate user-defined input parameters, while green, rectangular boxes represent calculations of respective quantities. An optional step (Step~0) allows deriving the current phase relation of a Josephson tunnel junction with inhomogeneous barrier thickness. Step~1 determines the current-flux relation of the rf-SQUID as a function of the externally applied magnetic flux. Step~2 calculates the resonance frequency shift using the current-flux relation obtained in step~1.}
    \label{fig:algorithm}
\end{figure}

Instead of increasing the order of the Taylor expansion of equation~\ref{eq:Is}, a more efficient and scalable approach is to solve equation~\ref{eq:Is} numerically. Figure~\ref{fig:algorithm} shows a flowchart of our algorithm for computing $I_\mathrm{S}(t)$ and determining the resulting resonance frequency shift. Assuming the current-phase relation $I_\mathrm{S,JJ}(\varphi)$ of the utilized Josephson junction, and thus the current-flux relation $I_\mathrm{S}(\varphi_\mathrm{tot})$ of the rf-SQUID, is known, the algorithm begins at step~1 by determining the time evolution of the supercurrent in the rf-SQUID as a function of the externally applied flux $\varphi_\mathrm{ext}$ for a given  flux amplitude of the microwave probe tone $\varphi_\mathrm{rf}$. This is accomplished using a fixed-point iteration method. The supercurrent is computed iteratively, where the $n$-th iteration is given by
\begin{equation}
    I_{\mathrm{S},n}(t)=I_\mathrm{S,JJ}(-\varphi_{\mathrm{tot},n}(t)). 
    \label{eq:I_S_num}
\end{equation}
Here, $\varphi_{\mathrm{tot},n}$ denotes the total normalized flux at the $n$-th iteration, calculated with the relation
\begin{equation}\varphi_{\mathrm{tot},n}(t)=\varphi_\mathrm{ext}+\varphi_\mathrm{rf}\sin{(\omega t)}+\beta_\mathrm{L}\frac{I_{\mathrm{S},n-1}(t)}{I_\mathrm{c}}.
\end{equation}
The iteration begins with a random guess $\varphi_\mathrm{tot,0}\in[0,2\pi)$ and the series is computed until it converges at iteration $n=N$, as defined by the stopping criterion
\begin{equation}
    |\varphi_{\mathrm{tot},N}-\varphi_{\mathrm{tot},N-1}|<\epsilon_\varphi,
\end{equation}
where $\epsilon_\varphi$ is a specified numerical tolerance. The final value $I_\mathrm{S}\equiv I_{\mathrm{S},N}$ then provides the desired supercurrent $I_\mathrm{S}(\varphi_\mathrm{ext},t)$. To reduce computation time, we apply acceleration techniques such as Aitken’s delta-squared process \cite{Saa25}. Our method reliably convergences for all tested rf-SQUID parameters. However, in the case of a hysteretic rf-SQUID (i.e., $\beta_\mathrm{L} > 1$), the result will depend on the initial guess $\varphi_{\mathrm{tot},0}$, and not all theoretically valid solutions can be found using this approach.

\begin{figure*}
    \centering
    \includegraphics[width=\linewidth]{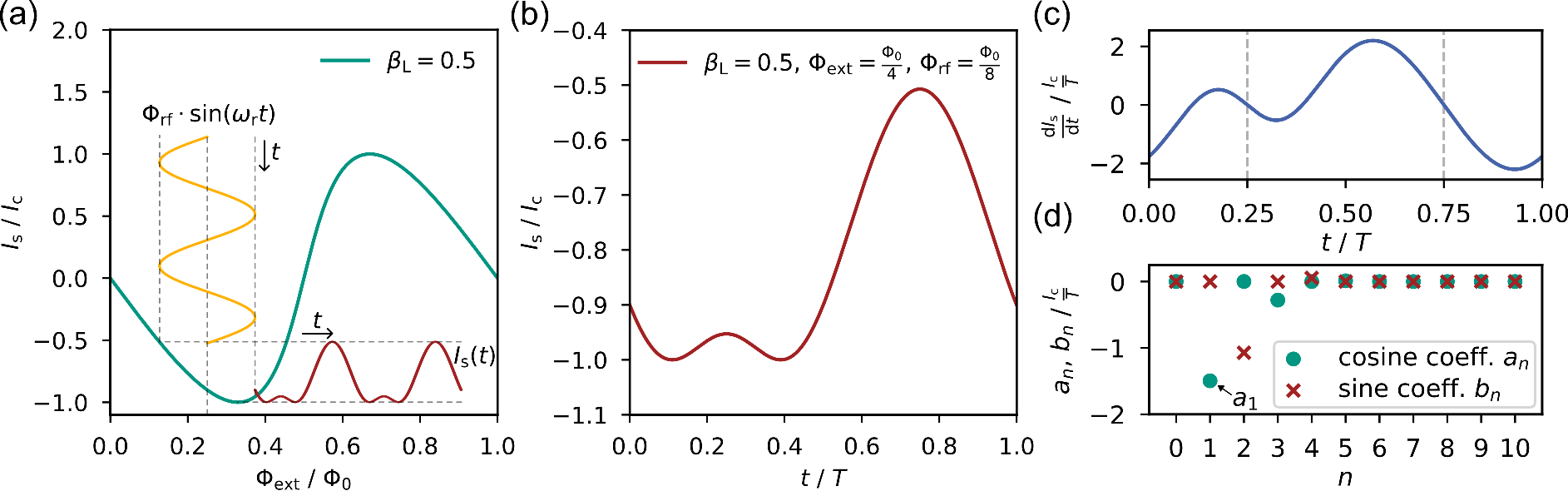}
    \caption{rf-SQUID supercurrent and its time derivative, calculated using our advanced $\upmu$MUX model, are shown. Panel (a) shows the SQUID's current–flux relation (green solid line). In addition, the calculated SQUID response (red solid line) to a microwave signal (yellow solid line) with amplitude $\Phi_\mathrm{rf} = \Phi_0/8$, applied at a flux offset $\Phi_\mathrm{ext} = \Phi_0/4$, is schematically depicted. Panel (b) and (c) show the time-dependence of the supercurrent $I_\mathrm{S}(t)$ and its time derivative, which result from the same microwave signal depicted in panel (a). The time derivative is anti-symmetric at $\pm T/4$ (gray dashed lines).  It contains multiple frequency components, which are extracted using a discrete Fourier transform. The corresponding cosine and sine coefficients $a_n$ and $b_n$ are shown in panel (d). When later calculating the effective load inductance $L_\mathrm{T,eff}$, only the first Fourier coefficient $a_1$ is relevant, as all other components either do not populate the resonator or vanish.}
    \label{fig:Is_to_Fourier}
\end{figure*}

Step~1 of our algorithm allows for the calculation of the SQUID supercurrent as a function of the externally applied flux, as shown in figure~\ref{fig:Is_to_Fourier}(a) for a representative SQUID with $\beta_\mathrm{L}=0.5$. To determine the induced current $i_\mathrm{ind}(t)$ in the load inductor, the time derivative $\mathrm{d}I_\mathrm{s}/\mathrm{d}t$ of $I_\mathrm{S}(t)$ is required. Therefore, we calculate the SQUID current as a function of time for one oscillation period $T=1/f_\mathrm{r}$, as depicted in figure~\ref{fig:Is_to_Fourier}(b), assuming a microwave probe tone with amplitude $\Phi_\mathrm{rf}=\Phi_0/8$ and an external flux bias $\Phi_\mathrm{ext}=\Phi_0/4$. 

Step~2 begins with computing the time derivative using a forward difference approximation:
\begin{equation}
    \frac{\mathrm{d}I_\mathrm{S}(t)}{\mathrm{d}t}\approx\frac{I_\mathrm{S}(t+\delta t)-I_\mathrm{S}(t)}{\delta t},
\end{equation}
where $\delta t\ll T$ is chosen to be sufficiently small.
Figure~\ref{fig:Is_to_Fourier}(c) shows $\mathrm{d}I_\mathrm{S}(t)/\mathrm{d}t$ over one oscillation period. It cannot be described by a simple sine function and therefore consists of multiple frequency components in addition to $\omega$. Consequently, according to equation~\ref{eq:ind}, we expect $i_\mathrm{ind}(t)$ to also consist of multiple frequency components. However, due to the resonance condition~\ref{eq:fr}, only the fundamental frequency $\omega\approx\omega_\mathrm{r}$ populates the resonator. Higher harmonics do not meet the resonance condition and are therefore suppressed by destructive interference.
Therefore, the time derivative $\mathrm{d}I_\mathrm{S}(t)/\mathrm{d}t$ is evaluated over one oscillation period $T$ and the $\omega$-component is determined using a discrete Fourier transform. Because of the anti-symmetry of $\mathrm{d}I_\mathrm{S}(t)/\mathrm{d}t$ at $t=T/4$ and $t=-T/4$\footnote{We define $\varphi_\mathrm{app}=\varphi_\mathrm{ext}+\varphi_\mathrm{rf}\sin{(\omega t)}$ as the flux applied to the rf-SQUID, which is symmetric around $t=\pm T/4$. By definition, for a non-hysteretic rf-SQUID, the implicit expression $\varphi_\mathrm{tot}=\varphi_\mathrm{app}+\beta_\mathrm{L}I_\mathrm{S}(\varphi_\mathrm{tot})/I_\mathrm{c}$ has a unique solution for each value of $\varphi_\mathrm{app}$. This means the mapping of $\varphi_\mathrm{app}$ to $\varphi_\mathrm{tot}$ is single-valued, and $\varphi_\mathrm{tot}$ is a well-defined function of $\varphi_\mathrm{app}$, which inherits the symmetry of $\varphi_\mathrm{app}$ around $t=\pm T/4$. Since $I_\mathrm{S}(\varphi_\mathrm{tot})=I_\mathrm{S,JJ}(-\varphi_\mathrm{tot})$ is also uniquely defined for any $\varphi_\mathrm{tot}$, $I_\mathrm{S}$ also inherits the symmetry around $t=\pm T/4$, making its time derivative $\mathrm{d}I_\mathrm{S}/\mathrm{d}t$ anti-symmetric around $t=\pm T/4$.}, it can be shown that the corresponding sine coefficient $b_1$ must be zero. For this reason, only the first cosine coefficient $a_1$ is needed to calculate $i_\mathrm{ind}(t)$ using equation~\ref{eq:ind}:
\begin{equation}
    i_\mathrm{ind}(t)=-\frac{M_\mathrm{T}}{\omega L_\mathrm{T}}a_1\sin{(\omega t)}.
\end{equation}
The induced current is phase shifted by $\pi/2$ due to the complex impedance $i\omega L_\mathrm{T}$ of the load inductor and therefore oscillates in phase with the current $i_\mathrm{T}(t)$. The resulting effective load inductance, calculated using equation~\ref{eq:LT}, is time-independent and can be used with equation~\ref{eq:fr} to determine the resonance frequency. Using this numerical approach, the $\upmu$MUX resonance shift can be evaluated as a function of $\varphi_\mathrm{ext}$ and $\varphi_\mathrm{rf}$ for any non-hysteretic rf-SQUID. Thus, under the assumption of a junction current-phase relation $I_\mathrm{S,JJ}(\varphi)=I_\mathrm{c}\sin{(\varphi)}$, any $\upmu$MUX with $\beta_\mathrm{L}<1$ can be modeled, making this model more general as compared to the analytical model described by Wegner {\it et al.}~\cite{Weg22}. 

\begin{figure*}
    \centering
    \includegraphics[width=\linewidth]{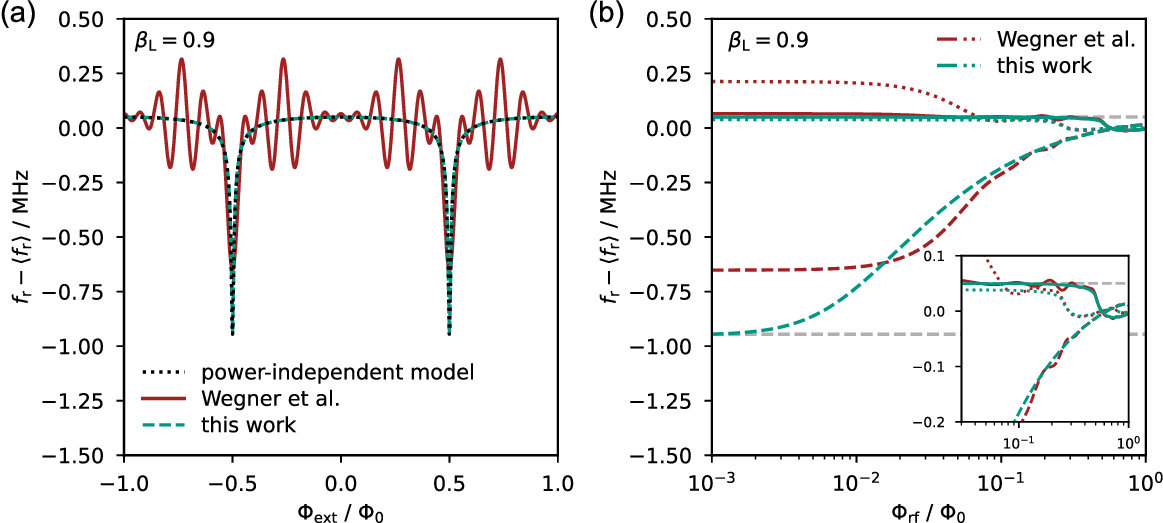}
    \caption{Comparison between the analytical $\upmu$MUX model presented by Wegner {\it et al.}~\cite{Weg22} and the present numerical model, assuming an rf-SQUID with hysteresis parameter $\beta_\mathrm{L} = 0.9$. Panel (a) displays the dependence of the resonance frequency $f_\mathrm{r}$ on the applied magnetic flux in the limit of negligible power of the microwave probe tone ($\Phi_\mathrm{rf} \rightarrow 0$). $\langle f_\mathrm{r} \rangle = 5\,\mathrm{GHz}$ denotes the resonance frequency averaged over $\Phi_\mathrm{ext}$. Panel (b) shows the resonance frequency as a function of microwave probe tone power $\Phi_\mathrm{rf}$. The solid green and red line corresponds to $\Phi_\mathrm{ext} = n\Phi_0$, the dashed green and red line to $\Phi_\mathrm{ext} = (n + 1/2)\Phi_0$, and the dotted green and red line to $\Phi_\mathrm{ext} = (n \pm 1/4)\Phi_0$. The dashed light gray lines mark the minimum and maximum resonance frequency predicted by the power-independent model. The numerical model shows the expected behavior and matches the exact power-independent solution in the low-power limit, while the analytical model exhibits ripples caused by the limitations of its Taylor expansion. The inset shows a zoom near $\Phi_\mathrm{rf} \approx 0.6\,\Phi_0$, where the resonance frequencies of the different flux states intersect for the first time.}
    \label{fig:shift_vs_power}
\end{figure*}

Figure~\ref{fig:shift_vs_power} compares both models, assuming a $\upmu$MUX channel with a screening parameter of $\beta_\mathrm{L}=0.9$. This parameter exceeds the applicable range of the analytical model presented by Wegner {\it et al.}~\cite{Weg22}, resulting in the significant ripples observed in figure~\ref{fig:shift_vs_power}(a). In contrast, our numerical model does not display these ripples and shows excellent agreement in the limit of low readout powers $\Phi_\mathrm{rf}\rightarrow 0$ with the exact but power-independent model described in \cite{Mat11,Kem17}.
When both models are evaluated for varying microwave probe tone power $\Phi_\mathrm{rf}$ (see figure~\ref{fig:shift_vs_power}(b)), our numerical model shows the expected decrease in the resonance frequency shift with increasing readout power. For screening parameters $\beta_\mathrm{L}<0.6$, we observe only negligible deviations between the resonance shift predicted by both models. However, for $\beta_\mathrm{L}>0.6$, the ripples in the analytical model result in significant deviations, with these deviations being more pronounced at lower readout powers. Consequently, only our model converges to the power-independent exact solution for $\Phi_\mathrm{rf}\rightarrow0$. 

A fit of the data shown in figure~\ref{fig:models_phirf0} using our numerical model shows significantly better agreement with the measured data (see figure~\ref{fig:models_phirf0}(b,e)) than the analytical model presented by Wegner {\it et al.}~\cite{Weg22}. This improvement becomes particularly evident in the residuals $r_i$, where the numerical model eliminates the unphysical ripples present in the analytical model. Moreover, the overall agreement between model and data, quantified by the fit’s coefficient of determination $R^2$ value, increases from $94.6\,\%$ to $95.5\,\%$.  However, the numerical model still underestimates the total resonance frequency shift and overestimates the resonance frequency between the modulation coil currents $I_\mathrm{mod}\approx0\,\mathrm{\upmu A}$ and $I_\mathrm{mod}\approx\,50\,\mathrm{\upmu A}$ corresponding to $\Phi_\mathrm{ext}\approx \pm\Phi_0/4$, suggesting an incomplete understanding of the behavior of this $\upmu$MUX channel. 

The current-phase relation $I_\mathrm{S,JJ}(\varphi)$ used in equation~\ref{eq:I_S_num} can be any well-defined, continuous function. This flexibility allows us to consider current–phase relations that deviate from the conventional sinusoidal form. Consequently, our numerical model can simulate $\upmu$MUX devices based on non-tunneling-type Josephson junctions, such as Dayem bridges or 3D nanobridges, simply by inserting the appropriate current–phase relation. The approach can also improve the description of devices based on Josephson tunnel junctions.
As shown in the following section, the fit to the data in figure~\ref{fig:models_phirf0} can be much improved by replacing the well-known sinusoidal current-phase relation of a Josephson tunnel junction with a more complex description, which accounts for inhomogeneities in the Josephson tunnel junction's barrier thickness.

% -----------------------------------------------

\section{Effects of Josephson tunnel junction barrier inhomogeneities on µMUX characteristics}

So far, we have used a sinusoidal current-phase relation $I_\mathrm{S,JJ} = I_\mathrm{c}\sin(\varphi)$, which assumes a homogeneous tunnel barrier and low transmission probability. However, in practice, the tunnel barrier of actual Josephson tunnel junctions displays a certain degree of roughness. Adam~\textit{et al.}, for instance, demonstrated that Nb/Al-AlO$_\mathrm{x}$/Nb based window-type junctions exhibit significant spatial variation in the critical current density across the cross-section of the tunnel barrier~\cite{Ada24}. Furthermore, Willsch~\textit{et al.} showed that superconducting qubits relying on Al/AlO$_\mathrm{x}$/Al Josephson tunnel junctions cannot be accurately modeled using a purely sinusoidal current-phase relation~\cite{Wil24}. Instead, they proposed a mesoscopic model of tunneling through an inhomogeneous AlO$_\mathrm{x}$ barrier. 

\begin{figure*}
    \centering
    \includegraphics[width=\linewidth]{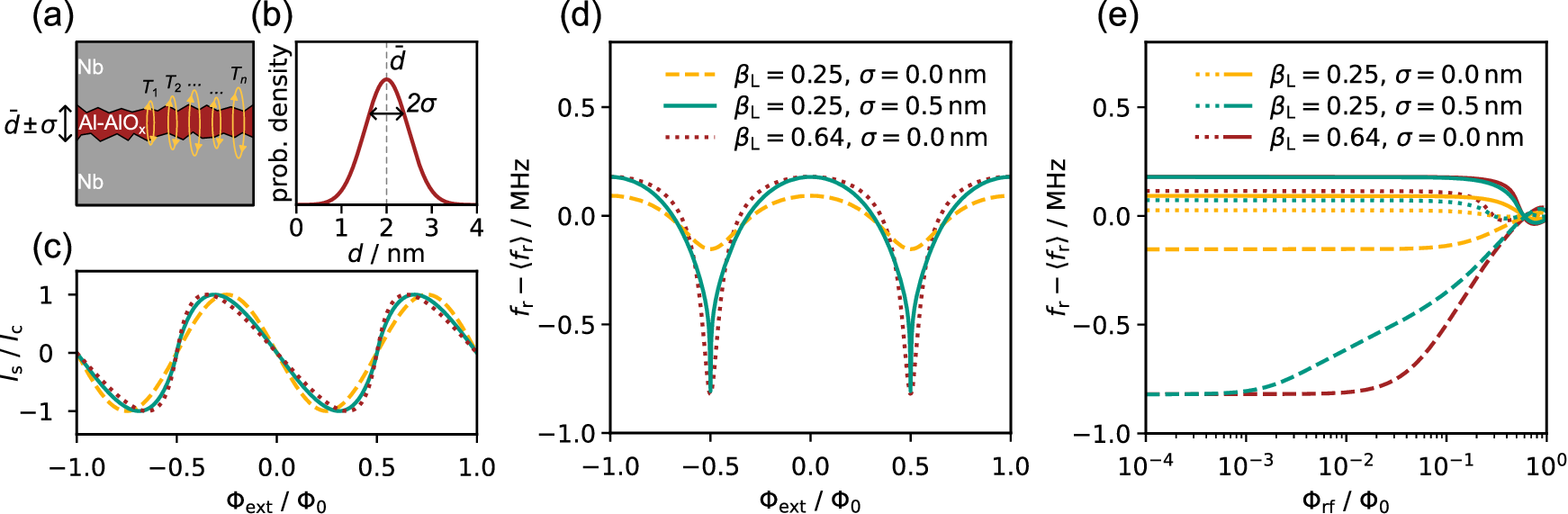}
    \caption{The figures show the effect of an inhomogeneous tunnel barrier with a mean thickness $\bar{d} = 2\,\mathrm{nm}$ and a standard deviation $\sigma = 0.5\,\mathrm{nm}$ on the $\upmu$MUX characteristics. Panel (a) schematically shows the cross section of a Nb/Al-AlO$_\mathrm{x}$/Nb Josephson tunnel junction with an inhomogeneous barrier thickness. As the barrier thickness varies along the junction, the transmission probabilities $T_1$ to $T_N$ of the $N$ individual conduction channels also vary. Panel (b) displays the assumed tunnel barrier thickness probability density distribution, a Gaussian distribution centered at the mean thickness $\bar{d}$. Panel (c) shows the current–flux relation of an rf-SQUID with $\beta_\mathrm{L} = 0.25$ for a homogeneous tunnel barrier (yellow, dashed line), and for an inhomogeneous barrier with $\sigma = 0.5\,\mathrm{nm}$ (green, solid line). For comparison, a homogeneous junction with a screening parameter $\beta_\mathrm{L} = 0.64$ is also shown (red, dotted line). Similar to an increased screening parameter, the barrier inhomogeneity causes the current–flux relation to appear skewed. Panel (d) shows the resonance frequency shift of $\upmu$MUX devices using the same SQUID parameters as in panel (c), with $\langle f_\mathrm{r} \rangle = 5\,\mathrm{GHz}$. Panel (e) presents the power dependence of the resonance frequency shift for external magnetic flux values $\Phi_\mathrm{ext} = n\Phi_0$ (solid line), $\Phi_\mathrm{ext} = (n + 1/2)\Phi_0$ (dashed line), and $\Phi_\mathrm{ext} = (n \pm 1/4)\Phi_0$ (dotted line). The effects of barrier inhomogeneity and increased screening parameter are qualitatively similar, however, they are distinct. This becomes evident in the significantly steeper resonance frequency shift of the inhomogeneous device near $(n + 1/2)\Phi_0$ in panel (d), as well as the reduced resonance frequency at $(n \pm 1/4)\Phi_0$, visible in panel (d) and, as a function of probe tone powers, in panel (e).}
    \label{fig:JJinhomo}
\end{figure*}

To incorporate inhomogeneous barriers into our advanced $\upmu$MUX model, the current-phase relation $I_\mathrm{S,JJ}(\varphi)$ used in step~1 in figure \ref{fig:algorithm} is replaced by a relation directly derived from the mesoscopic model. This requires the introduction of an additional calculation step (step~0 in figure \ref{fig:algorithm}), in which the current-phase relation is derived using the mesoscopic model proposed by Willsch~\textit{et al.}~\cite{Wil24}. The model assumes a rough tunneling barrier between the two junction electrodes, as illustrated in figure \ref{fig:JJinhomo}(a). Locally, the barrier is described as a rectangular potential of thickness $d$ and height $\phi$. The local thickness $d$ follows a Gaussian distribution with mean $\bar{d}$ and standard deviation $\sigma$, as depicted in figure~\ref{fig:JJinhomo}(b). Within this model, assuming operation well below the critical temperature, the current-phase relation of the Josephson junction is expressed as a Fourier series in terms of the Josephson energies $E_{\mathrm{J},k}$:
\begin{equation}
    I_\mathrm{S,JJ}(\varphi) = \frac{1}{\Phi_0} \sum_{k=1}^{\infty} k E_{\mathrm{J},k} \sin(k\varphi).
    \label{eq:Is_inhomo}
\end{equation}
The Josephson energies depend on the junction parameters and can be expressed as
\begin{equation}
    E_{\mathrm{J},k} = \frac{\Delta}{4} \frac{N}{k} \int_0^1 \mathrm{d}T\, \rho(T) c_k(T).
\end{equation}
where $\Delta$ denotes the superconducting energy gap of the junction electrodes and $N$ the number of conduction channels across the junction barrier. The Fourier coefficients $c_k(T)$ depend solely on the transmission probability $T$ and are independent of the parameters specific to the junction
\begin{equation}
    c_k(T)=\left( \begin{array}{c}
    2k-2 \\
    k-1
    \end{array} \right) \frac{(-1)^{k+1}T^k}{16^{k-1}}
    \cdot {}_2F_1\left(k-\frac{1}{2}, k+\frac{1}{2};2k+1;T\right), \nonumber
\end{equation}
where ${}_2F_1$ denotes the hypergeometric function.

The distribution of transmission probabilities $T$ across the conduction channels is described by the function $\rho(T)$. For the mesoscopic model with Gaussian barrier thickness distribution, the distribution $\rho(T)$ is given by~\cite{Wil24}
\begin{equation}
    \rho(T) = \frac{2}{1 + \mathrm{Erf}(\tilde{d} / \sqrt{2} \tilde{\sigma})} \frac{1}{2T \sqrt{1 - T + a^2 T}}
    \cdot\frac{1}{\sqrt{2\pi} \tilde{\sigma}} \exp{\left({-\frac{(f(T) + \alpha)^2}{2 \tilde{\sigma}^2}}\right)},\nonumber
\end{equation}
where $\mathrm{Erf}(x)$ is the error function, $\tilde{d}=\bar{d}/d_0$, $\tilde{\sigma}=\sigma/d_0$, and

\begin{equation}
    d_0 = \hbar/\sqrt{2m_\mathrm{i}\phi},
\end{equation}
\begin{equation}
    a^2=\frac{1}{4}\left( \sqrt{\frac{m_\mathrm{i}\phi}{m_e E_\mathrm{F}}}-\sqrt{\frac{m_e E_\mathrm{F}}{m_\mathrm{i}\phi}} \right)^2,
\end{equation}
\begin{equation}
    f(T)=\log{\left( \frac{\sqrt{T}}{\sqrt{1-T}+\sqrt{1-T+a^2T}}\right)},
\end{equation}
\begin{equation}
    \alpha=\tilde{d}+\log{(a)}.
\end{equation}
Here, $E_\mathrm{F}$ denotes the Fermi energy of the junction electrodes, $m_e$ the mass of the free electron, and $m_i$ the effective electron mass in the band of the junction barrier closest to the energy of the tunneling electrons.

Since the total number of conduction channels $N$ is generally unknown, the absolute values of the Josephson energies $E_{\mathrm{J},k}$ cannot be determined. However, the relative magnitudes of the Josephson energies are sufficient to reconstruct the current-phase relation up to a scaling factor. The scaling factor can be determined by normalizing $I_\mathrm{S,JJ}(\varphi)$ so that its maximum corresponds to the critical current $I_\mathrm{c}$. 

The Josephson energies $E_{\mathrm{J},k}$ alternate in sign and decrease exponentially with increasing $k$. The algorithm computes these energies up to $k=K$, such that
\begin{equation}
    \left|\frac{E_{\mathrm{J},K}}{E_{\mathrm{J},1}}\right|<\epsilon_E
\end{equation}
where $\epsilon_E\ll1$ represents a sufficiently small numerical threshold. For larger inhomogeneities $\sigma$, the Josephson energies decrease at a slower rate, which necessitates the inclusion of terms with higher $k$. In contrast, for homogeneous barriers, it is sufficient to consider only the first Josephson energy, simplifying equation~\ref{eq:Is_inhomo} to the well-known sinusoidal current-phase relation.

To apply this model to our $\upmu$MUX devices, which are based on Nb/Al-AlO$_\mathrm{x}$/Nb Josephson tunnel junctions, we assume a typical barrier thickness of $\bar{d}=2\,\mathrm{nm}$, a barrier height of $\phi=1\,\mathrm{eV}$, and an effective electron mass of $m_\mathrm{i}=0.75m_e$~\cite{Jun09,Rip02}. For the Fermi energy of niobium, we use $E_\mathrm{F}=5.32\,\mathrm{eV}$~\cite{Ash76}. The numerical tolerances were set to $\epsilon_\varphi=10^{-10}$, $\delta t =10^{-6}T$, and $\epsilon_E=10^{-10}$. 

Figure \ref{fig:JJinhomo}(c) shows the current-flux relation of an rf-SQUID, calculated using our numerical model. We compare the cases of a homogeneous and an inhomogeneous tunnel barrier and also examine the effect of an increased screening parameter. The inhomogeneity is set to $\sigma=0.5\,\mathrm{nm}$, leading to significant higher-order contributions from the Josephson energies ($|E_\mathrm{J,2}/E_\mathrm{J,1}|\approx3.9\,\%$, $|E_\mathrm{J,3}/E_\mathrm{J,1}|\approx0.9\,\%$, ...). As a result, the current-flux relation of the SQUID appears skewed, resembling the effect caused by a large screening parameter. The curve becomes steeper around $\Phi_\mathrm{ext}=(n+1/2)\Phi_0$ and shallower around $\Phi_\mathrm{ext}=n\Phi_0$. Consequently, the total resonance shift (see figure \ref{fig:JJinhomo}(d)) increases and becomes more asymmetric around the average resonance frequency $\langle f_\mathrm{r}\rangle$.
While the inhomogeneity can produce the same total resonance frequency shift around $f_0$ as the screening parameter, the overall shape of this shift is distinct. This becomes particularly evident when comparing the power dependence of both cases (see figure \ref{fig:JJinhomo}(e)). For the inhomogeneous tunnel barrier, the resonance frequency at $\Phi_\mathrm{ext}=(n\pm1/4)\Phi_0$ is lower, and the reduction in total resonance frequency shift begins at lower readout powers $\Phi_\mathrm{rf}$. Because the effects of junction inhomogeneity and large screening parameters are qualitatively similar, neglecting inhomogeneity during device characterization could result in an incorrect estimate of the device's screening parameter. 

\begin{figure*}
    \centering
    \includegraphics[width=\linewidth]{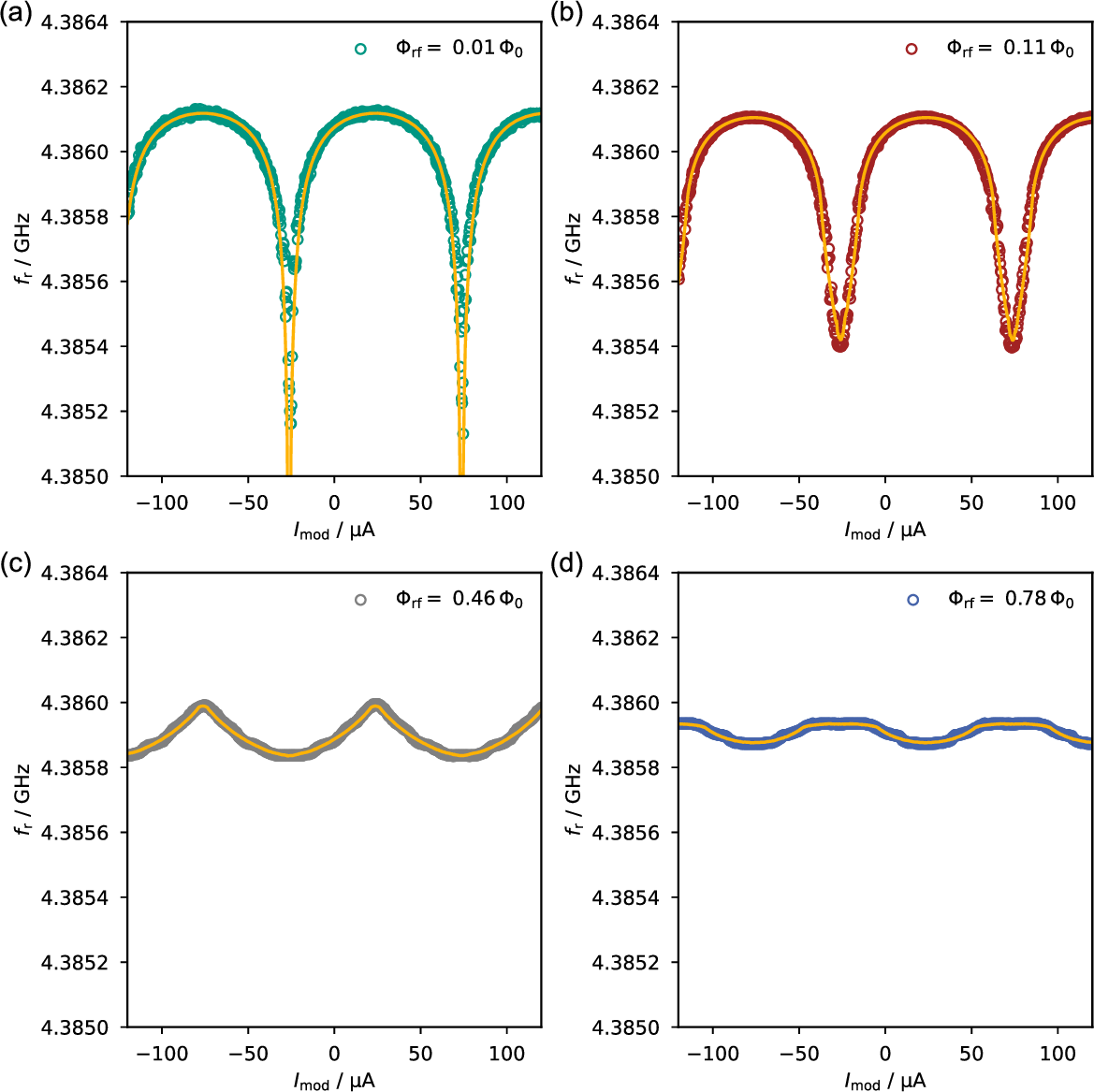}
    \caption{Dependence of the measured resonance frequency $f_\mathrm{r}(I_\mathrm{mod})$ on the modulation coil current $I_\mathrm{mod}$ for microwave probe tone powers ranging from $\Phi_\mathrm{rf} = 0.01\,\Phi_0$ to $\Phi_\mathrm{rf} = 0.78\,\Phi_0$. The yellow lines represent the numerical model assuming a junction inhomogeneity of $\sigma = 0.48\,\mathrm{nm}$, showing excellent agreement with the measured data.}
    \label{fig:shift_multiple_Pwrs}
\end{figure*}

We fitted our numerical model incorporating an inhomogeneous tunnel barrier to the data presented in figure \ref{fig:models_phirf0}(c)(f). For this, we again assume a mean barrier thickness of $\bar{d}=2\,\mathrm{nm}$, a barrier height of $\phi=1\,\mathrm{eV}$, an effective electron mass of $m_\mathrm{i}=0.75\,m_\mathrm{e}$ in the barrier, and a niobium Fermi energy of $E_\mathrm{F}=5.32\,\mathrm{eV}$. Only the inhomogeneity $\sigma$ is added as additional free fit parameter. The best agreement between the data and the model was achieved with an inhomogeneity of $\sigma=0.48\,\mathrm{nm}$, representing a notable improvement over the homogeneous barrier model. The $R^2$ value is improved from $95.5\,\%$ to $97.4\,\%$ and the slight curvature of the residuals $r_i$ observed in the homogeneous model is reduced. The same model applied to data taken from the same device at higher readout powers $\Phi_\mathrm{rf}$, is shown in figure \ref{fig:shift_multiple_Pwrs}. We observe excellent agreement between the model and measurement even at $\Phi_\mathrm{rf}=0.78\,\Phi_0$, which exceeds the typically applied readout powers. This demonstrates that our advanced model remains accurate across the full range of screening parameters and readout powers relevant for $\upmu$MUX operation.

Although the measured data strongly suggest the presence of inhomogeneities in tunnel barrier thickness in the measured $\upmu$MUX device, it is important to note that the exact distribution of barrier thickness, as well as the exact tunneling parameters $E_\mathrm{F}$, $m_\mathrm{i}$, and $\Phi$ remain unknown. The critical current densities reported in \cite{Ada24} show spatial variations that deviate from the simple Gaussian thickness distribution assumed in our Gaussian approach. Additionally, fabrication-related defects may create high-transparency conduction channels that are not accounted for by the Gaussian model. However, such deviations from the assumed transmission probability distribution $\rho(T)$ are expected to produce qualitatively similar effects. Consequently, minor discrepancies between the model and the experimental data can likely be attributed to uncertainties in the actual form of $\rho(T)$.

\section{Conclusion}
We presented a numerical model for the readout power dependence of the $\upmu$MUX resonator characteristics. The model is valid for SQUID screening parameters up to $\beta_\mathrm{L}<1$, covering the full range of practically relevant design parameters. We demonstrated that, for devices with $\beta_\mathrm{L} > 0.6$, the numerical model significantly improves agreement with experimental data compared to existing models, thereby enabling optimization beyond the previously accessible parameter space. Moreover, the current-phase relation of the Josephson junction assumed in the model can be easily modified, allowing for the inclusion of $\upmu$MUX devices based on non-tunneling type Josephson junctions or devices based on Josephson tunnel junctions with inhomogeneous tunnel barriers. We have shown that the effects of such inhomogeneities are qualitatively similar to, yet distinct from, those of the screening parameter, making their inclusion essential for accurate characterization. Incorporating these effects yields improved agreement with measurements, even at readout powers well beyond typical operating conditions. 
Thus, our model provides a powerful tool for the design and analysis of $\upmu$MUX devices, well-suited for integration into simulation frameworks aimed at optimizing next-generation $\upmu$MUX readout systems.

\ack{
M. Neidig gratefully acknowledges support from the Karlsruhe School of Elementary Particle and Astroparticle Physics: Science and Technology (KSETA). This work was partially funded by the Deutsche Forschungsgemeinschaft (DFG, German Research Foundation) – Projektnummer (project number) 467785074.}

\section*{References}
\bibliographystyle{iopart-num}
\bibliography{literature.bib}
\end{document}